\documentclass[preprint, 10pt, 5p]{elsarticle}
\usepackage[utf8]{inputenc}
\usepackage{geometry} 
\usepackage{graphicx} 
\usepackage{amsmath, amsbsy}
\usepackage{amssymb}
\usepackage{float} 
\usepackage{wrapfig} 
\usepackage[hidelinks]{hyperref}
\usepackage{tikz}

\biboptions{numbers, sort&compress}

\newcommand{\bv}{\mathbf{v}}
\newcommand{\br}{\mathbf{r}}
\newcommand{\bA}{\mathbf{A}}
\newcommand{\bB}{\mathbf{B}}
\newcommand{\bE}{\mathbf{E}}
\newcommand{\bu}{\mathbf{u}}

\newcommand{\D}{\text{d}}

\newcommand{\bell}{\ensuremath{\boldsymbol\ell}}

\newcommand{\blambda}{\ensuremath{\boldsymbol{\lambda}}}
\newcommand{\bbeta}{\ensuremath{\boldsymbol{\beta}}}

\begin{document}
\title{Maximum-Entropy States for Magnetized Ion Transport}
\date{\today}
\author{E. J. Kolmes}
\ead{ekolmes@princeton.edu}
\author{I. E. Ochs}
\author{M. E. Mlodik}
\author{N. J. Fisch}
\address{Department of Astrophysical Sciences, Princeton University, Princeton, NJ 08544, USA}

\begin{abstract}
For a plasma with fixed total energy, number of particles, and momentum, the distribution function that maximizes entropy is a Boltzmann distribution. 
If, in addition, the rearrangement of charge is constrained, as happens on ion-ion collisional timescales for cross-field multiple-species transport, the maximum-entropy state is instead given by the classic impurity pinch relation. The maximum-entropy derivation, unlike previous approaches, does not rely on the details of the collision operator or the dynamics of the system, only on the presence of certain conservation properties. 
\end{abstract}

\begin{keyword}
Maximum entropy \sep plasma thermodynamics \sep differential transport \sep impurity transport \sep impurity pinch \sep rotating plasma 
\end{keyword}

\maketitle

\section{Introduction}

Cross-field particle transport is a problem of central importance throughout much of plasma science. Many controlled fusion technologies, including magnetic mirrors \cite{Post1987, Fetterman2008, Gueroult2012}, tokamaks \cite{Hirshman1981, Redi1991, Fisch1992, Wade2000, Dux2004, Hole2014, Zonca2015}, and stellarators \cite{McCormick2002, Braun2010, Helander2017, Newton2017}, rely on the physics of cross-field transport.
Other applications include devices such as plasma mass filters \cite{Bonnevier1966, Lehnert1971, Hellsten1977, Krishnan1983, Geva1984, Prasad1987, Bittencourt1987, Grossman1991, Fetterman2011b, Ochs2017iii, Dolgolenko2017, Zweben2018, Gueroult2018}, magnetic traps for high-$Z$ ion sources \cite{Geller1985, Geller1986}, plasma thrusters \cite{Smirnov2004, Ellison2012}, and non-neutral particle traps \cite{Davidson1970, ONeil1979, Prasad1979, ONeil1981, Imajo1997, Dubin1999}. 
Similar concerns (though involving quite different parameter regimes) appear in plasma astrophysics \cite{Jokipii1966, Goldreich1969, Qin2002, Kulsrud}. 

The relative motion of different ion species is often of particular significance. 
In a fusion plasma, differential cross-field transport is important for fuel injection and ash removal. In any hot plasma or implosion experiment, high-$Z$ impurities have the potential to radiate a great deal of power, so purging them can be important. On the other hand, for certain radiation sources in the x-ray regime, the higher-$Z$ ions are useful \cite{Ryutov2000, Bailey2004, Rochau2008}, and understanding their behavior is important. For a number of diagnostic applications using trace impurity radiation \cite{Rochau2008, Stratton2008}, the degree to which impurities concentrate must be determined in order to inform correctly on the ambient plasma. 
Moreover, differential cross-field ion transport is crucially important for plasma mass filters, which are designed to separate out different components of a plasma. 

In a fully ionized, magnetized plasma, in the absence of temperature gradients, it is a well-known result of cross-field transport theory that collisional species $a$ and $b$ reach steady state only when \cite{Spitzer1952, Taylor1961ii, Braginskii1965, Connor1973, Rutherford1974, Hinton1974, Ochs2017}
\begin{gather}
n_a^{1/Z_a} \propto n_b^{1/Z_b} . \label{eqn:impurityPinch}
\end{gather}
In the presence of a species-dependent potential $\Phi_s$, this becomes \cite{Kolmes2018}
\begin{gather}
\big( n_a e^{\Phi_a/T} \, \big)^{1/Z_a} \propto \big( n_b e^{\Phi_b/T} \, \big)^{1/Z_b} . \label{eqn:generalizedPinch}
\end{gather}
Eq.~(\ref{eqn:impurityPinch}), which holds for any number of species, is what we might call the classic impurity pinch relation. 
Eqs.~(\ref{eqn:impurityPinch}) and (\ref{eqn:generalizedPinch}), and special cases thereof, appear in descriptions of differential ion transport for a variety of different applications \cite{Spitzer1952, Taylor1961ii, Braginskii1965, Bonnevier1966, Connor1973, Rutherford1974, Hinton1974, Hinton1974, Hirshman1981, ONeil1981, Krishnan1983, Geva1984, Prasad1987, Bittencourt1987, Grossman1991, Imajo1997, Dubin1999, HelanderSigmar, Dolgolenko2017, Ochs2017, Kolmes2018, Ochs2018i, Ochs2018ii}. For instance, the case where $\Phi_s$ is the centrifugal potential is centrally important in descriptions of species separation in rotating plasmas. 

These equations have been derived using a number of different approaches and in a number of different contexts. 
Previous treatments proceed by writing down a model for the dynamics of the system, then studying the requirements for the system to reach steady state. Such an approach inevitably brings with it a number of assumptions: for instance, choices about the form of the collision operator and the parameter regime. 

These assumptions are often natural and reasonable, but they raise a question: what characteristics of a system are necessary in order to produce Eqs.~(\ref{eqn:impurityPinch}) and (\ref{eqn:generalizedPinch})? 
The answer is not obvious; the usual approach does not readily yield generalizations. 

This paper generalizes the previous derivations of Eqs.~(\ref{eqn:impurityPinch}) and (\ref{eqn:generalizedPinch}) by deriving them from a thermodynamic perspective. This new derivation provides physical insight into their origins. 
Section~\ref{sec:boltzmann} shows how the calculus of variations can be used to directly maximize the entropy in a cylinder of plasma with some fixed energy, particle number, and momentum and some externally imposed magnetic field; the result is the Boltzmann distribution, up to a frame transformation. 
Section~\ref{sec:pinch} describes how an additional constraint on charge transport leads instead to Eqs.~(\ref{eqn:impurityPinch}) and (\ref{eqn:generalizedPinch}). Section~\ref{sec:discussion} discusses the context and implications of this result. 

\section{Boltzmann Distribution from Maximum Entropy}\label{sec:boltzmann}

Consider a fully ionized, ideal plasma with $N$ species, containing $N_s$ particles of species $s$. Suppose the joint distribution can be written as 
\begin{gather}
f(\{\br_{sj}, \bv_{sj}\}) = \prod_{s=1}^{N} \prod_{j=1}^{N_s} \frac{f_s(\br_{sj},\bv_{sj})}{N_s} , \label{eqn:jointProbability}
\end{gather}
such that $f_s(\br,\bv) \, \D^3 \br \, \D^3 \bv / N_s$ is the probability of finding a given particle of species $s$ in an infinitesimal phase space volume $\D^3 \br \, \D^3 \bv$. 
The product of the one-particle distribution functions is typically a good approximation of the joint distribution when the number of particles in a Debye sphere is large \cite{Ichimaru}. 
For non-Cartesian coordinates, take the expression $\D^3 \br \, \D^3 \bv$ to include the appropriate Jacobian determinant implicitly (for instance, in spherical coordinates, take the convention that $\D^3 \br = r^2 \sin \theta \, \D r \, \D \theta \, \D \phi$). 

The entropy can be written up to multiplicative and additive constants as \cite{Jaynes1963}
\begin{gather}
S = - \int \bigg( \prod_{s=1}^N \prod_{j=1}^{N_s} \D^3 \br_{sj} \, \D^3 \bv_{sj} \bigg) f \log f \label{eqn:generalEntropy} .
\end{gather}
Eq.~(\ref{eqn:jointProbability}) can be combined with Eq.~(\ref{eqn:generalEntropy}) to get 
\begin{gather}
S = - \sum_{s=1}^N \int \D^3 \br \, \D^3 \bv \, f_s \log \bigg( \frac{f_s}{N_s} \bigg). 
\end{gather}

Suppose the system has a fixed number of particles $N_s$ for each species and a fixed total energy $\mathcal{E}$. These constraints can be written in terms of $f_s$ as 
\begin{gather}
N_s = \int \D^3 \br \, \D^3 \bv \, f_s \label{eqn:Ns}
\end{gather}
and
\begin{gather}
\mathcal{E} = \int \D^3 \br \bigg[ u_\text{EM} + \sum_s \int \D^3 \bv \, \bigg( \frac{1}{2} m_s |\bv|^2 + \Phi_s \bigg) f_s \bigg] , \label{eqn:E}
\end{gather}
where $u_\text{EM}$ is the energy density in the electromagnetic fields and $\Phi_s(\br)$ is any externally imposed non-electromagnetic potential that does not depend on $f_s$ (such as an imposed gravitational potential). 

Depending on the geometry of the system, there may also be some conserved momenta. The linear momentum could be written as 
\begin{gather}
\mathbf{P} = \int \D^3 \br \, \bigg( \mathbf{p}_\text{EM} + \sum_s \int \D^3 \bv \, m_s \bv f_s \bigg). \label{eqn:totalMomentum}
\end{gather}
In a system with appropriate rotational symmetries, there can also be a conserved angular momentum 
\begin{gather}
\mathbf{L} = \int \D^3 \br \, \bigg( \bell_\text{EM} + \sum_s \int \D^3 \bv \, m_s \br \times \bv f_s \bigg). \label{eqn:totalAngularMomentum}
\end{gather}
Here $\mathbf{p}_\text{EM}$ and $\bell_\text{EM}$ are the linear and angular momentum densities in the electromagnetic fields. 

In order to compute the choices of $f_s$ that extremize the entropy $S$ while conserving $N_s$, $\mathcal{E}$, and $\mathbf{L}$, consider a small perturbation to $f_s$ such that $f_s \rightarrow f_s + \delta f_s$. Let $\delta X / \delta f_s$ denote the functional derivative of $X$ with respect to $f_s$; if $X$ is expressed as a $k$-dimensional integral, 
\begin{gather}
\frac{\D X(f_s + \epsilon \delta f_s)}{\D \epsilon} \bigg|_{\epsilon=0} = \int \D^k \mathbf{y} \, \frac{\delta X}{\delta f_s} \, \delta f_s .
\end{gather}
Then the condition for an entropy-extremizing $f_s$ is 
\begin{align}
&\frac{\delta S}{\delta f_s} = \lambda_{N_s} \frac{\delta N_s}{\delta f_s} + \lambda_\mathcal{E} \frac{\delta \mathcal{E}}{\delta f_s} \nonumber \\
&\hspace{80 pt} + \blambda_\mathbf{P} \cdot \frac{\delta \mathbf{P}}{\delta f_s}  + \blambda_\mathbf{L} \cdot \frac{\delta \mathbf{L}}{\delta f_s} \, , \label{eqn:extremalCondition}
\end{align}
where $\lambda_{N_s}$ and $\lambda_\mathcal{E}$ are scalar constants and $\blambda_\mathbf{P}$ and $\blambda_\mathbf{L}$ are, in general, vector constants. 
Of course, for $s' \neq s$, $\delta N_{s'} / \delta f_s = 0$. 

The derivatives of $S$ and $N_s$ have the simplest dependences on $f_s$ and are straightforward to obtain. The entropy derivative is 
\begin{gather}
\frac{\delta S}{\delta f_s} = - \log \bigg( \frac{f_s}{N_s} \bigg) - 1
\end{gather}
and the particle number derivative is 
\begin{gather}
\frac{\delta N_s}{\delta f_s} = 1. 
\end{gather}
The other functional derivatives are a little more involved because the fields depend on the behavior of the charged particles. 

\subsection{Geometry, Boundary Conditions, and Fields} \label{subsec:geometry}

Consider a cylindrical $(r, \theta, z)$ geometry with unit vectors $\hat r$, $\hat \theta$, and $\hat z$. In order to externally impose a magnetic field, surround the plasma with a superconducting boundary at $r = R$; this will confine the plasma to $r < R$ and exclude $\bE$ and $\bB$ fields from $r > R$. 
Assume homogeneity in the $\hat \theta$ and $\hat z$ directions. 
In order to keep the constraints finite, assume that they are all taken over a region of some axial length $h$. 

The calculations that follow will neglect the effects of fluctuations in the plasma. This is typically a safe approximation. For instance, in a macroscopically neutral plasma the energy due to fluctuations will be smaller than the thermal energy by a factor of $\mathcal{O}(\Lambda^{-1})$, where $\Lambda$ is the plasma parameter \cite{KrallTrivelpiece}. 

Suppose the plasma is net neutral, so that although there may be places where $\rho_c(\br) \neq 0$ locally, there do not need to be surface charges on the boundary in order to exclude the field from $r > R$. Then the electric field is given by 
\begin{gather}
\bE(r) = \frac{\hat r}{\epsilon_0 r} \int_0^r \D r' \, r' \rho_c(r')
\end{gather}
and the electric potential is 
\begin{align}
\varphi(r) &= \frac{1}{\epsilon_0} \int_0^r \D r' \, r' \log \bigg( \frac{r'}{r} \bigg) \rho_c(r') . 
\end{align}

The boundary condition will enforce the conservation of magnetic flux, which is equivalent to fixing the cross-section-averaged field 
\begin{gather}
B_0 \doteq \frac{2}{R^2} \int_0^R \D r' \, r' B_z(r') .
\end{gather}
Let $\bB = \bB_\text{p} + \bB_\text{ext}$, where $\bB_\text{p}$ is the field generated by any currents in the plasma and $\bB_\text{ext}$ is the field generated by the external superconductor. The current density can be defined by 
\begin{gather}
\mathbf{j}(r) \doteq \int \D^3 \bv \, q_s \bv f_s(r,\bv) . \label{eqn:currentDensity}
\end{gather}
Then 
\begin{align}
\bB_\text{p}(r) &= \mu_0 \bigg[ \hat z \int_r^R \D r' j_\theta(r') \nonumber \\
&\hspace{50 pt}+ \frac{\hat \theta}{r} \int_0^r \D r' \, r' j_z(r') \bigg] 
\end{align}
and $\bB_\text{ext} = B_\text{ext} \hat z$ is described by 
\begin{gather}
B_\text{ext} = B_0 - \frac{2}{R^2} \int_0^R \D r' \, r' \hat z \cdot \bB_\text{p}(r') ,
\end{gather}
which is 
\begin{gather}
B_\text{ext} = B_0 - \mu_0 \int_0^R \D r' \, \bigg( \frac{r'}{R} \bigg)^2 j_\theta(r'). 
\end{gather}
Fixing a (Lorenz) gauge, the corresponding vector potential components can be written as 
\begin{align}
A_{\theta} &= \frac{\mu_0}{2r} \bigg[ \int_0^r \D r' \, (r')^2 j_\theta(r') + r^2 \int_r^R \D r' \, j_\theta(r') \nonumber \\
&\hspace{30 pt}- \frac{r^2}{R^2} \int_0^R \D r' \, (r')^2 j_\theta(r') \bigg] + \frac{r B_0}{2} \label{eqn:ATheta}
\end{align}
and
\begin{align}
A_z &= \mu_0 \int_r^R \frac{\D r'}{r'} \int_0^{r'} \D r'' \, r'' j_z(r'') . \label{eqn:AZ}
\end{align}

This choice of geometry and boundary conditions is intended to be the simplest one that illustrates all of the relevant physics. A cylinder has translational and rotational symmetry, so it can have conserved linear and angular momenta. The choice of superconducting boundaries allows a magnetic field to be imposed on the system. This will be important in Section~\ref{sec:pinch}, which discusses additional constraints characteristic of magnetized systems. 

These fields could also be imposed with external coils held at some constant current. However, this would introduce some complications. For instance, if the plasma were perturbed in a way that changed the flux inside the coils, it would be necessary to account for the energy expended in order to keep the coil current constant. The superconducting boundaries result in cleaner conservation laws. 

\subsection{Momentum}

The total linear and angular momenta are given by Eq.~(\ref{eqn:totalMomentum}) and Eq.~(\ref{eqn:totalAngularMomentum}), respectively. In a cylinder, set $\mathbf{P} = P \hat z$ and $\mathbf{L} = L \hat z$. The momentum density in the electromagnetic fields is 
\begin{gather}
\mathbf{p}_\text{EM} = \epsilon_0 \bE \times \bB
\end{gather}
and the angular momentum in the fields is 
\begin{gather}
\mathbf{\bell}_\text{EM} = \epsilon_0 \br \times (\bE \times \bB). 
\end{gather}
Using these expressions, together with the description in Section~\ref{subsec:geometry} of the dependence of $\bE$ and $\bB$ on $f_s$, it is possible to show that the functional derivative of the linear momentum is 
\begin{gather}
\frac{\delta P}{\delta f_s} = \bigg( m_s \bv + q_s \bA + \frac{q_s \bv \varphi}{c^2} \bigg) \cdot \hat z
\end{gather}
and the functional derivative of the angular momentum is 
\begin{align}
\frac{\delta L}{\delta f_s} &= m_s r v_\theta + q_s r A_\theta - \frac{q_s R^2 B_0}{2} \nonumber \\
&\hspace{10 pt} - \frac{q_s v_\theta}{c^2} \frac{1}{r} \int_0^r \D r' \, (r')^2 E_r(r') \nonumber \\
&\hspace{10 pt} + \frac{q_s v_\theta}{c^2} \frac{r}{R^2} \int_0^R \D r' \, (r')^2 E_r(r') . 
\end{align}

\subsection{Energy}

The field energy density $u_\text{EM}$ can be written as 
\begin{gather}
u_\text{EM} = \frac{\epsilon_0 E^2}{2} + \frac{B^2}{2 \mu_0} \, .
\end{gather}
Using this, let $\mathcal{E} = \mathcal{E}_\text{p} + \mathcal{E}_\text{EM}$, where 
\begin{gather}
\mathcal{E}_\text{p} = \int \D^3 \br \, \D^3 \bv \, \bigg( \frac{1}{2} m_s | \bv |^2 + \Phi_s \bigg) f_s \\
\mathcal{E}_\text{EM} = \int \D^3 \br \, \bigg( \frac{\epsilon_0 E^2}{2} + \frac{B^2}{2 \mu_0} \bigg) . 
\end{gather}
These variations can be taken separately. Keeping in mind that $\Phi_s$ does not depend on $f_s$, 
\begin{gather}
\frac{\delta \mathcal{E}_\text{p}}{\delta f_s} = \frac{1}{2} m_s | \bv |^2 + \Phi_s . 
\end{gather}
The part due to the fields is, after some calculation, 
\begin{gather}
\frac{\delta \mathcal{E}_\text{EM}}{\delta f_s} 
= q_s \varphi + q_s \bv \cdot \bA - \frac{q_s r v_\theta B_0}{2} \, .
\end{gather}
This $\bA$ includes both the plasma-generated field and the one imposed by the boundary conditions. 

\subsection{Constructing a Distribution Function}

The distribution function $f_s(r,\bv)$ can now be calculated by inserting the functional derivatives into Eq.~(\ref{eqn:extremalCondition}). Without simplification, this results in a fairly unwieldy expression. Relabel the Lagrange multipliers by defining the following constants: 
\begin{align}
T &\doteq \frac{1}{\lambda_\mathcal{E}} \\
u_z &\doteq - T \blambda_\mathbf{P} \cdot \hat z \\
\Omega &\doteq - T \blambda_\mathbf{L} \cdot \hat z \\
f_{s0} &\doteq N_s e^{-1-\lambda_{N_s} - q_s R^2 \Omega B_0/2T + m_s u_z^2 / 2T} .
\end{align}
Define the velocity $\bu \doteq u_z \hat z + r \Omega \, \hat \theta$. In the non-inertial frame moving at $\bu$, to leading order in $u/c$, the electric potential is \cite{Jackson, Thyagaraja2009} 
\begin{gather}
\tilde \varphi \doteq \varphi - \bu \cdot \bA .
\end{gather}
This is the same as the relativistic expression with the Lorentz factor $\gamma$ set to 1; recall that the original expressions for the system's conserved quantities were also written in their nonrelativistic forms. 

In the frame moving at $\bu$, the externally imposed potential includes a centrifugal part, so that 
\begin{gather}
\tilde{\Phi}_s \doteq \Phi_s - \frac{m_s r^2 \Omega^2}{2} \, .
\end{gather}
Finally, define a modified vector potential $\tilde{\bA}$ by 
\begin{align}
\tilde{A}_\theta &\doteq A_\theta - \frac{r B_0}{2} + \frac{\Omega}{r c^2} \int_0^r \D r' (r')^2 E_r(r') \nonumber \\
&\hspace{50 pt} - \frac{r \Omega}{R^2 c^2} \int_0^R \D r' (r')^2 E_r(r') \label{eqn:AThetaHat}
\end{align}
and 
\begin{align}
\tilde{A}_z &\doteq A_z - \frac{u_z \varphi}{c^2} \, .\label{eqn:AZFrame}
\end{align}
$\tilde{A}_\theta$ includes frame-transformation terms as well as terms due to the externally imposed field. 
$\tilde{A}_z$ is the $\hat z$ component of the vector potential in the moving frame (with $\gamma \rightarrow 1$). Strictly speaking, whether $u_z \varphi / c^2$ is first or second order in $u/c$ depends on the relative ordering of $u$ and $\varphi/A_z$. In either case, Eq.~(\ref{eqn:AZFrame}) is accurate up to a relativistic $\mathcal{O}(u^2/c^2)$ correction. 

Using these definitions, 
\begin{align}
f_s &= f_{s0} \exp \bigg[ - \frac{m_s (\bv - \bu + q_s \tilde{\bA} / m_s)^2}{2 T} - \frac{q_s \tilde{\varphi}}{T} \nonumber \\
&\hspace{20 pt} - \frac{\tilde{\Phi}_s}{T} + \frac{m_s (\bu - q_s \tilde{\bA}/m_s)^2 - \bu^2}{2T} \bigg]. \label{eqn:distributionWithA}
\end{align}
In the frame moving at $\bu$, species $s$ will have an average velocity of $-q_s \tilde{\bA} / m_s$. Recall that the gauge for the vector potential was fixed in Eqs.~(\ref{eqn:ATheta}) and (\ref{eqn:AZ}). It follows immediately that the only self-consistent choice for the $\hat z$ component is $\tilde{A}_z = 0$, since a nonzero value would need to be generated by currents moving in one direction but would produce currents that move in the opposite direction. 

In order to understand the $\hat \theta$ component, note that Eq.~(\ref{eqn:AThetaHat}) can also be written as 
\begin{align}
\tilde{A}_\theta = \frac{1}{r} \bigg[ \int_0^r \D r' \, r' \tilde{B}_z - \frac{r^2}{R^2} \int_0^R \D r' \, r' \tilde{B}_z \bigg], 
\end{align}
where 
\begin{align}
\tilde{B}_z(r') \doteq B_z(r') + \frac{r' \Omega E_r(r')}{c^2} 
\end{align}
is $B_z$ in the moving frame. In other words, $\tilde{A}_\theta$ is positive at $r$ if the region within $r$ has an average $\tilde{B}_z$ that is larger than the average within $R$ and negative in the opposite case. Amp\`{e}re's law specifies that the current in the moving frame is $\tilde{j}_\theta = - \partial \tilde{B}_z / \partial r$. The sign of $\tilde{j}_\theta$ is opposite the sign of $\tilde{A}_\theta$, so if at some $r$ the average field within $r$ is larger than that within $R$, $\tilde{B}_z$ will be increasing at that $r$; in the opposite case, it will be decreasing. This implies that the average enclosed field at $r'$ must be above (or below) that within $R$ for all $r' > r$, which leads to a contradiction at $r'=R$. As such, $\tilde{A}_\theta$ must vanish everywhere. 

In the end, the self-consistency of $\tilde{\bA}$ requires 
\begin{align}
f_s &= f_{s0} \exp \bigg[ - \frac{m_s (\bv - \bu)^2}{2 T} - \frac{q_s \tilde{\varphi}}{T} - \frac{\tilde{\Phi}_s}{T} \bigg] . \label{eqn:boltzmannDistribution}
\end{align}
The expression for $\tilde{\varphi}$ retains a dependence on the magnetic field, in spite of the conventional intuition that thermodynamic equilibria should not have such a dependence. However, in the (generally non-inertial) frame that is comoving with the plasma, this dependence disappears and $f_s$ is the expected Boltzmann distribution. To see this, note that $\bu$ is the bulk velocity of the plasma, and recall that $\tilde{\varphi}$ and $\tilde{\Phi}_s$ are the potentials evaluated in the frame moving at $\bu$. 
The Boltzmann distribution for a plasma, as well as generalizations that include conserved momenta, has been derived from constrained maximum-entropy techniques elsewhere \cite{Stankiewicz1968, Garrett1988, Garrett1990i, Garrett1990ii, Dubin1999}, though the superconducting boundary conditions and self-consistent treatment of the self-organized electromagnetic fields used here are unusual. A closely related discussion, albeit for a system with somewhat different boundary conditions, can be found in Dubin \cite{Dubin1999}. 

Eq.~(\ref{eqn:boltzmannDistribution}) specifies the form of the maximum-entropy state in terms of the parameters $T$, $u_z$, $\Omega$, and $f_{s0}$. In order to fully determine $f_s$ for a particular choice of the global constraints $N_s$, $\mathcal{E}$, $P_z$, and $L_z$, it would be necessary to map these constraints to the four parameters in the solution. 
This mapping is described implicitly by Eqs.~(\ref{eqn:Ns}), (\ref{eqn:E}), (\ref{eqn:totalMomentum}), (\ref{eqn:totalAngularMomentum}), and (\ref{eqn:boltzmannDistribution}). 
Problems of this kind are not trivial \cite{Davidson1970, ONeil1979, Prasad1979, ONeil1981, Garrett1988, Garrett1990i, Garrett1990ii, Dubin1999}, though in some cases it is possible to read off self-consistent solutions. If $P_z = L_z = 0$, if there is no additional imposed $\Phi_s$, and if $\sum_s q_s N_s = 0$, it is consistent to pick $u_z = \Omega = 0$, $f_{s0} = N_s (m_s / 2 \pi T)^{3/2} / \pi R^2 h$, and $T = (\mathcal{E} - \pi R^2 h B_0^2 / 2 \mu_0) / \sum_s (3 N_s / 2)$. Computing general, explicit expressions for the constants $T$, $u_z$, $\Omega$, and $f_{s0}$ is not the focus of this paper. In the general case, one would have to both find consistent choices of the parameters and determine their uniqueness. 

Even without explicit expressions for $T$, $u_z$, $\Omega$, and $f_{s0}$, Eq.~(\ref{eqn:boltzmannDistribution}) contains information about the maximum-entropy states of the plasma. For instance, the net flow of the plasma must be a superposition of solid-body rotation and axial translation. The flow must be the same for all species; there can be currents, but only as a result of flow in regions where there is charge imbalance. In any case, for the analysis in Section~\ref{sec:pinch}, the actual values of the parameters in Eq.~(\ref{eqn:boltzmannDistribution}) are not needed.

\section{Constrained Charge Transport}\label{sec:pinch}

This section will use the formal machinery and intuition from Section~\ref{sec:boltzmann} to derive Eqs.~(\ref{eqn:impurityPinch}) and (\ref{eqn:generalizedPinch}). Physically, the difference between Eq.~(\ref{eqn:generalizedPinch}) and Eq.~(\ref{eqn:boltzmannDistribution}) comes from the way in which magnetic fields restrict plasma transport. 

For motivation, consider the case of magnetized cross-field transport in a system containing two species $a$ and $b$ with densities $n_a$ and $n_b$. If the interaction between the two species produces a cross-field force $\mathbf{F}_{ab}$ on species $a$ and $\mathbf{F}_{ba}$ on species $b$, then $n_a \mathbf{F}_{ab} = - n_b \mathbf{F}_{ba}$. The resulting cross-field fluxes will be 
\begin{gather}
\Gamma_{ab} = \frac{n_a \mathbf{F}_{ab} \times \hat b}{q_a B} = - \frac{q_b}{q_a} \, \Gamma_{ba}, 
\end{gather}
where $\hat b \doteq \mathbf{B} / B$. 
If these are the dominant cross-field fluxes, then it follows from the continuity equation that $\partial_t(q_a n_a + q_b n_b) \approx 0$, so the local charge density is approximately fixed; there may be other processes that can push net charge across the field lines, but they are typically slow. 

With that in mind, consider a charge density constraint: 
\begin{gather}
\psi(\br, f_{s \notin \mathcal{C}}) = \int \D^3 \bv \sum_{s \in \mathcal{C}} q_s f_s(\br,\bv) . \label{eqn:chargeConstraint}
\end{gather}
This is the local charge density due to the species in some set $\mathcal{C}$. The distinction between $s \in \mathcal{C}$ and $s \notin \mathcal{C}$ accommodates the possibility that some species may be able to move across field lines more easily than others. 
For instance, it is often appropriate over ion-ion collisional timescales not to include electrons in $\mathcal{C}$, since the classical cross-field transport associated with electron-ion collisions is much slower than the transport associated with ion-ion collisions; the accumulation of ions generally associated with Eq.~(\ref{eqn:impurityPinch}) takes place on this faster ion-ion timescale. Indeed, if Eq.~(\ref{eqn:impurityPinch}) is applied to all ion and electron species in a quasineutral plasma, it requires that all density profiles be flat. It might also be appropriate to include additional constraints involving only $s \notin \mathcal{C}$ (i.e., fixing the electron density to its initial profile), but this would not affect the rest of the analysis. 

Similarly, an arbitrary dependence of $\psi$ on $f_{s \notin \mathcal{C}}$ is allowed but not required. 
The constraint on $\psi$ can be used to fix the charge density to its initial value, but note that formally it does not have to do so. Also note that Eq.~(\ref{eqn:chargeConstraint}) was motivated by the vanishing net flux of charge, and that it would lead to a constraint of that form, but that the constraint could be imposed for any other reason with the same effect. 

$\psi(\br)$ is different from the constraints introduced in Section~\ref{sec:boltzmann} in that it specifies a value for every point in space rather than a single scalar or vector for the whole system. Instead of a constant Lagrange multiplier, the condition for the extremal $f_s$ pairs $\delta \psi(\br) / \delta f_s$ with a multiplier that is a function of $\br$:
\begin{align}
\frac{\delta S}{\delta f_s} &= \lambda_{N_s} \frac{\delta N_s}{\delta f_s} + \lambda_{\mathcal{E}} \frac{\delta \mathcal{E}}{\delta f_s} 
\nonumber \\ &\hspace{10 pt}
+ \blambda_\mathbf{P} \cdot \frac{\delta \mathbf{P}}{\delta f_s}
+ \blambda_\mathbf{L} \cdot \frac{\delta \mathbf{L}}{\delta f_s}
+ \lambda_\psi (\br) \frac{\delta \psi}{\delta f_s} \, . \label{eqn:pinchExtremalCondition}
\end{align}
For any $s \in \mathcal{C}$, 
\begin{gather}
\frac{\delta \psi}{\delta f_s} = q_s. 
\end{gather}
Continuing to use the definitions from Section~\ref{sec:boltzmann} for $\bu$, $\tilde{\bA}$, $\tilde{\varphi}$, and $f_{s0}$, the resulting distribution is 
\begin{align}
f_s &= f_{s0} \exp \bigg[ - \frac{m_s (\bv - \bu)^2}{2 T} - \frac{\tilde{\Phi}_s}{T} \nonumber \\
&\hspace{90 pt} - q_s \bigg( \lambda_\psi(\br) + \frac{\tilde{\varphi}}{T} \bigg) \bigg] . \label{eqn:fPinch}
\end{align}
Recall that $f_{s0}$, $\mathbf{u}$, and $T$ do not depend on $\br$. Even without determining $\lambda_\psi(\br)$, it follows that 
\begin{align}
&\bigg\{ \frac{f_a}{f_{a0}} \exp \bigg[ \frac{m_a (\bv-\mathbf{u})^2}{2T} + \frac{\tilde{\Phi}_a}{T} \bigg] \bigg\}^{1/Z_a} \nonumber \\
&\hspace{10 pt}= \bigg\{ \frac{f_b}{f_{b0}} \exp \bigg[ \frac{m_b (\bv-\bu)^2}{2T} + \frac{\tilde{\Phi}_b}{T} \bigg] \bigg\}^{1/Z_b}
\end{align}
for any $a,b \in \mathcal{C}$. Here $Z_s \doteq q_s / e$. 

Integrating out the velocity dependence gives back a familiar expression: 
\begin{gather}
\big( n_a e^{\tilde{\Phi}_a/T} \, \big)^{1/Z_a} \propto \big( n_b e^{\tilde{\Phi}_b/T} \, \big)^{1/Z_b} . \label{eqn:derivedGeneralizedPinch}
\end{gather}
The difference between this and Eq.~(\ref{eqn:generalizedPinch}) is a matter of whether or not $\Phi_s$ is defined from the beginning to include effective potential terms like the centrifugal potential. The expression takes the same form whether or not $\Phi_s$ is defined to include the electrostatic potential, since $\varphi_s \propto Z_s$ and therefore cancels. In the case where either $\tilde{\Phi}_s = 0$ or $\tilde{\Phi}_s \propto Z_s$, Eq.~(\ref{eqn:derivedGeneralizedPinch}) reduces to Eq.~(\ref{eqn:impurityPinch}). Note that Maxwell-Boltzmann distributions satisfy Eq.~(\ref{eqn:derivedGeneralizedPinch}), but that they are a special case; Eq.~(\ref{eqn:derivedGeneralizedPinch}) holds even when Eq.~(\ref{eqn:boltzmannDistribution}) does not. 

The same physics that is responsible for the conservation of $\psi$ may be associated with additional physical constraints, even if they are not necessary in order to derive Eq.~(\ref{eqn:derivedGeneralizedPinch}). For instance, in the Braginskii fluid model \cite{Braginskii1965}, viscous interactions are one of the major mechanisms that can move net charge across field lines. The smallness of the viscous force compared to the local friction between ion species is one of the reasons why $\psi$ can be treated as constant over sufficiently short timescales. The viscous force is also one of the major mechanisms for the spatial redistribution of mechanical momentum. If the viscosity is approximated as small, it may also be appropriate to enforce local momentum conservation laws, such as 
\begin{gather}
\bell(\br) = \int \D^3 \bv \sum_{s \in \mathcal{C}} m_s \br \times \bv f_s, 
\end{gather}
or more generally, constraints of the form 
\begin{gather}
\xi(\br) = \int \D^3 \bv \sum_{s \in \mathcal{C}} \bv \cdot \bbeta_s(\br) f_s. 
\end{gather}
Constraints of this type would change the velocity-space structure of Eq.~(\ref{eqn:fPinch}) without changing Eq.~(\ref{eqn:derivedGeneralizedPinch}), so there is no physical requirement that Eq.~(\ref{eqn:derivedGeneralizedPinch}) must be associated with solid-body rotation or uniform axial translation. 

\section{Discussion}\label{sec:discussion}

In the existing literature, Eqs.~(\ref{eqn:impurityPinch}) and (\ref{eqn:generalizedPinch}) have been derived from fluid models \cite{Spitzer1952, Rutherford1974, Kolmes2018}, using a jump-moment formalism \cite{Taylor1961ii}, by solving kinetic equations \cite{Connor1973}, and from a single-particle perspective \cite{Ochs2017}. In all of these cases, the resulting expression is understood as a condition for the steady state of some particular model for the time-evolution of the plasma. 

Although these derivations all end with Eqs.~(\ref{eqn:impurityPinch}) or (\ref{eqn:generalizedPinch}), they leave open questions about the general class of collision operators that will lead to the same result. 
The calculation in Section~\ref{sec:pinch} gives a different (and perhaps more fundamental) derivation of Eqs.~(\ref{eqn:impurityPinch}) and (\ref{eqn:generalizedPinch}). 
This calculation does not rely on the details of the form of the collision operator. In order to produce the classic impurity accumulation result, it is sufficient for a system to ($i$) conserve energy, ($ii$) constrain the motion of charge across field lines, and ($iii$) cause the system to attain its maximum-entropy state subject to ($i$) and ($ii$). 

The derivation in Section~\ref{sec:pinch} also includes conservation laws for particle number and momentum. These are useful for understanding the system, but they are not necessary in order to derive Eq.~(\ref{eqn:derivedGeneralizedPinch}) (though the inclusion of the constraint on $\mathbf{L}$ changes the rotation profile, which affects the effective potential). Non-conservation of $N_s$, $\mathbf{L}$, and $\mathbf{P}$ simply results in a particular choice of $f_{s0}$ and in $u_z = \Omega = 0$. 

There are ways in which the thermodynamic derivation of Eqs.~(\ref{eqn:impurityPinch}) and (\ref{eqn:generalizedPinch}) is more general than previous derivations, but there is an important way in which it is less general: in the form presented here, it does not capture the effects of temperature gradients or of temperature differences between species. The maximum-entropy state naturally enforces a spatially uniform temperature that is the same for all species. The physics of differential ion transport in the presence of temperature gradients and temperature differences is of significant theoretical and practical interest \cite{Taylor1961ii, Rutherford1974, Wade2000, Dux2004, Kagan2012, Kagan2014, Ochs2018i, Ochs2018ii}. 

Of course, in a real system, the charge density constraint described by Eq.~(\ref{eqn:chargeConstraint}) is not exact. Even in a strongly magnetized, quiescent plasma, there are a variety of mechanisms that can drive cross-field currents. These include viscous forces, inertial effects, and collisions with neutral particles \cite{HelanderSigmar, Rozhansky2008, Rax2019, Kolmes2019}. The important thing is that these processes be comparatively slow, so that the system will smoothly transition between states that satisfy Eq.~(\ref{eqn:generalizedPinch}) as $\psi$ varies. Approximate conservation laws of this kind have historically played an important role in the theoretical understanding of plasma relaxation processes \cite{Chew1956, Woltjer1958, Taylor1974, Taylor1986, Helander2017ii, Kaur2018, Kaur2019}. 

\section*{Acknowledgements}
This work was supported by NSF PHY-1506122, DOE DE-FG02-97ER25308, and NNSA 83228-10966 [Prime No. DOE (NNSA) DE-NA0003764].

\bibliographystyle{apsrev4-1} 
\bibliography{variationalBib.bib}

\end{document}